\documentclass[preprint,preprintnumbers,amsmath,amssymb]{revtex4}

\usepackage{bm}
\usepackage{graphicx}
\usepackage{subfig}
\usepackage{array}
\usepackage{color}

\graphicspath{{figures/}}

\makeatletter

\newcommand{\Rmnum}[1]{\expandafter\@slowromancap\romannumeral #1@}
\makeatother

\begin{document}

\title{Tests of Lorentz and CPT Violation in the Medium Baseline Reactor Antineutrino Experiment}

\author{Yu-Feng Li}\email{liyufeng@ihep.ac.cn}\quad \author{Zhen-hua Zhao}\email{zhaozhenhua@ihep.ac.cn}
\affiliation{Institute of High Energy Physics, Chinese Academy of Sciences,\\
P.O. Box 918, Beijing 100049, China}

\vspace{1.5cm}

\begin{abstract}
Tests of Lorentz and CPT violation in the medium baseline reactor antineutrino experiment are presented
in the framework of the Standard Model Extension (SME). Both the spectral distortion and sidereal variation are
employed to derive the limits of Lorentz violation (LV) coefficients. We do the numerical analysis of the sensitivity
of LV coefficients by taking the Jiangmen Underground Neutrino Observatory (JUNO) as an illustration,
which can improve the sensitivity by more than two orders of magnitude compared with the current limits
from reactor antineutrino experiments.
\end{abstract}


\maketitle

\section{Introduction}

Special Relativity (SR) is a fundamental theory describing the Lorentz
space-time symmetry, which is a consequence of the homogeneous and
isotropic space-time and the principle of relativity among different
inertial frames. Being a cornerstone of modern physics, SR has been
verified with an extremely high degree of
accuracy \cite{test}. Although Lorentz invariance has been widely accepted,
well-motivated models with Lorentz violation (LV) are anticipated from the
principle of Quantum Gravity \cite{ssb}. LV can be tested with the observations of high energy
cosmic rays \cite{review}, astrophysical neutrinos \cite{uhenu} and gamma-ray bursts \cite{grb},
as well as the low energy phenomena such as
beta decays \cite{beta}, double beta decays \cite{betabeta}, atomic co-magnetometer experiments \cite{atomic}
and neutrino oscillations \cite{nulv}.

The phenomena of neutrino oscillations have been well established from recent oscillation experiments with solar neutrinos,
atmospheric neutrinos, reactor antineutrinos and accelerator neutrinos \cite{PDG}.
The standard picture of three active neutrino oscillations is described by three
mixing angles, two independent mass-squared differences
and the CP-violating phase. The next-generation neutrino oscillation experiments with higher precision are designed
to probe the neutrino mass ordering and CP violation, and they also provide an opportunity to test new physics \cite{NP} beyond the Standard Model (SM).

The low energy phenomena of LV can be systematically studied in the framework of the
Standard Model Extension (SME) \cite{sme}, which includes all possible LV terms formed by
the SM fields in the Lagrangian,
\begin{equation}
 {\mathcal L}_{LV} \hspace{0.2cm}\simeq - (a_{L})_\mu \overline{\psi_{L}} \gamma^\mu \psi_{L}
 -(c_{L})_{\mu \nu} \overline{\psi_{L}} i \gamma^\mu \partial^\nu \psi_{L}.\label{eq:lagrangian}
\end{equation}
Notice that $a_{L}$ violates both the Lorentz and CPT symmetries,
but $c_{L}$ is CPT-even and only violates the Lorentz invariance.
Although the LV coefficients are very small due to the suppression factor of the order of the electro-week scale divided by the Planck scale (i.e, $10^{-17}$),
they provide an accessible test to Planck scale physics.

The SME framework predicts distinct behaviors for neutrino flavor conversions,
which are very different from the standard picture of three active neutrino oscillations. The transition probability depends on the ratio
of neutrino propagation distance $L$ and the neutrino energy $E$ (i.e., $L/E$) in
the conventional oscillation theory, but in the SME it depends on either $L$ or $L\times E$ for the contribution induced by $a_{L}$ or $c_{L}$.
On the other hand, LV also predicts the breakdown of space-time's isotropy, which manifests as a sidereal modulation
of the neutrino events for experiments with both the neutrino source and detector fixed on the earth. The sidereal time is
defined on the basis of the earth's orientation with respect to the sun-centered reference inertial frame, which will be defined in the
next section. The sidereal variation of the experimental event rate
will signify unambiguous evidence for LV. Previous LV searches in LSND \cite{lsnd}, MINOS \cite{MINOS_near,MINOS_far},
MiniBooNE \cite{MiniBooNE} and Double Chooz \cite{chooz} are reported with the limit level ranging from $10^{-20}$ to $10^{-23}$,
where $a_{L}$ is in the unit of GeV and $c_{L}$ is unitless.

In the present work we will study the possible tests of Lorentz and CPT violation in the medium baseline reactor antineutrino experiment.
We reformulate the theoretical description of vacuum neutrino oscillations in the SME framework, and discuss the key factors that affect
the sensitivity of LV. Using both the spectral distortion and sidereal variation effects,
we present the sensitivity of LV coefficients for the medium baseline reactor antineutrino experiment
taking the Jiangmen Underground Neutrino Observatory (JUNO) as an illustration.
Our analyses are carried out in the reference frame of the sun,
which is defined in Ref.~\cite{frame}, and has been used in the neutrino oscillation experiments \cite{lsnd,MINOS_near,MINOS_far,MiniBooNE,chooz}.

The remaining part of this work is organized as follows. Section 2
is to derive the theoretical description. We present the neutrino oscillation probability in the SME framework,
and define the standard and local frames to denote the direction-dependent property of LV.
In section 3, we give the numerical analysis for JUNO and calculate its sensitivity to the relevant LV coefficients. Finally, we conclude in section 4.

\section{Theoretical Description}
According to the SME, the effective Hamiltonian for Lorentz violating neutrino oscillations is written as  \cite{lv}
\begin{equation}
 H_{\alpha \beta}=\displaystyle \frac{1}{E}\hspace{0.05cm}\left[\hspace{0.05cm}\displaystyle \frac{m^2}{2}+(a_{L})_\mu
 p^{\mu}+(c_{L})_{\mu\nu} p^{\mu}p^{\nu}\hspace{0.05cm}\right]_{\alpha \beta}, \label{eq:hamitonian}
\end{equation}
where $E$ and $p^{\mu}$ are the neutrino energy and 4-momentum,  $\alpha$ and $\beta$ are flavor indices, and $(m^2)_{\alpha \beta}=U^{}{\rm diag}(m^{2}_{1},m^{2}_{2},m^{2}_{3})U^{\rm \dagger}$ is the mass-squared matrix in the flavor basis with $U$ being
the Pontecorvo-Maki-Nakagawa-Sakata (PMNS) matrix \cite{PMNS}. $a_{L}$ and $c_{L}$ are the LV coefficients defined in
Eq.~(\ref{eq:lagrangian}), and from dimensional analyses one can observe that $a_{L}$ is dimension-one while $c_{L}$ is dimensionless.
Taking account of the CPT transformation property of $H$ \cite{cpt}, we find that the Hamiltonian for antineutrinos has the following form
\begin{equation}\label{eq:anhamitonian}
 H_{\alpha \beta}=\displaystyle \frac{1}{E}\hspace{0.05cm}\left[\hspace{0.05cm}\displaystyle \frac{m^2}{2}-(a_{L})_\mu
 p^{\mu}+(c_{L})_{\mu\nu} p^{\mu}p^{\nu}\hspace{0.05cm}\right]_{ \alpha \beta}^{*}\,.
\end{equation}
To derive the antineutrino oscillation probabilities from the above Hamiltonian,
we employ the perturbative treatment as in Ref.~\cite{long} to factorize the LV part from the
conventional neutrino oscillation part. This approximate treatment is reasonable
for the situation with significant oscillation effects (i.e., long baseline).

In the standard framework of three active antineutrino oscillations, the oscillation probability
can be expressed as the square of a time evolution operator $S^{(0)}$:
\begin{equation}\label{p0v}
P^{(0)}_{\alpha\rightarrow\beta}=\left|S^{(0)}_{\alpha \beta}\right|^2=\left|(e^{-i H_0 t})_{\alpha \beta}\right|^2=
\left|\sum_i \hspace{0.05cm}U_{\alpha i}\hspace{0.05cm} U_{\beta i}^*\hspace{0.05cm}e^{-i E_i t}\right|^2,
\end{equation}
where $H_0=(m^2)^*/(2E)$ is the mass term of the Hamiltonian in Eq.~(\ref{eq:anhamitonian}), and
$t\simeq L$ for ultra-relativistic particles.
By defining $\delta H = [-(a_{L})_\mu p^{\mu}+(c_{L})_{\mu\nu} p^{\mu}p^{\nu}]^{*}/E$ and taking the perturbative expansion,
we derive the whole time evolution operator as
\begin{equation}\label{sfull}
\begin{array}{ll}
\vspace{0.2cm}
 S&=\hspace{0.2cm} e^{-iH t}\hspace{0.2cm}= \hspace{0.2cm}\left(e^{-i H_0 t-i\delta H t}\hspace{0.05cm}e^{i H_0 t}\right)\hspace{0.05cm}e^{-iH_0t}\\
 \vspace{0.2cm}
 &=\hspace{0.2cm}\left(1-i\int_0^t dt_1 \hspace{0.05cm} e^{-i H_0 t_1}\hspace{0.05cm}\delta H \hspace{0.05cm} e^{i H_0 t_1}+\cdots\right)S^{(0)}\\ &=\hspace{0.2cm}S^{(0)}+S^{(1)}+\cdots,
 \end{array}
\end{equation}
where the higher order contributions are omitted. Consequently, the oscillation probability is expanded as
\begin{equation}\label{pab}
 P_{\alpha\rightarrow\beta}\hspace{0.1cm}=\hspace{0.1cm}\left|S_{\alpha \beta}\right|^2\hspace{0.1cm}=\hspace{0.1cm}\left|\hspace{0.05cm}S^{(0)}_{\alpha \beta}\hspace{0.05cm}\right|^2+2\hspace{0.05cm}{\rm Re}\hspace{0.05cm}\left[\hspace{0.05cm}(S^{(0)}_{\alpha \beta})^*\hspace{0.05cm}S^{(1)}_{\alpha \beta}\hspace{0.05cm}\right]\hspace{0.1cm}=\hspace{0.1cm} P^{(0)}_{\alpha\rightarrow\beta}+P^{(1)}_{\alpha\rightarrow\beta}.
\end{equation}
In particular, the explicit form of $P^{(1)}_{\alpha\rightarrow\beta}$ can be written as \cite{long}
\begin{equation}\label{p1}
P^{(1)}_{\alpha\rightarrow\beta}=\displaystyle{\sum_{i,j}} \hspace{0.05cm} {\displaystyle\sum_{\rho,\sigma}}\hspace{0.05cm}2\hspace{0.05cm} L\hspace{0.05cm} {\rm Im}\hspace{0.05cm}\left[\hspace{0.05cm}(S^{(0)}_{\alpha \beta})^*\hspace{0.05cm}U_{\alpha i}\hspace{0.05cm} U_{\rho i}^* \hspace{0.05cm}\tau_{ij} \hspace{0.05cm}\delta H_{\rho \sigma}\hspace{0.05cm} U_{\sigma j} \hspace{0.05cm}U_{\beta j}^*\hspace{0.05cm}\right],
\end{equation}
with
\begin{equation}\label{tauij}
  \tau_{ij}=\left \{
\begin{array}{cc}
\vspace{0.2cm}
\exp{\{-iE_i L\}}&\hspace{0.3cm} {\rm when}\hspace{0.3cm}i=j\\
\displaystyle \frac{\exp{\{-iE_i L\}}-\exp{\{-iE_j L\}}}{-i(E_i-E_j)L}&\hspace{0.3cm} {\rm when}\hspace{0.3cm}i\neq j
\end{array}
\right..
\end{equation}

The hermiticity of Hamiltonian requires $\delta H_{\rho \sigma}$ to be real when $\rho =\sigma$
and $\delta H_{\rho \sigma}=\delta H_{ \sigma \rho}^*$ when $\rho \neq \sigma$.
For the former case, the LV terms can be extracted directly as
\begin{equation}\label{p1eq}
P^{(1)}_{\alpha\rightarrow\beta}=\displaystyle{\sum_{i,j}}\hspace{0.05cm}{\displaystyle\sum_{\rho=\sigma}}\hspace{0.05cm} 2\hspace{0.05cm}L\hspace{0.05cm} {\rm Im}\hspace{0.05cm}\left[(S^{(0)}_{\alpha \beta})^*\hspace{0.05cm}U_{\alpha i}\hspace{0.05cm} U_{\rho i}^* \hspace{0.05cm}\tau_{ij} \hspace{0.05cm}\hspace{0.05cm} U_{\sigma j} \hspace{0.05cm}U_{\beta j}^*\right]\delta H_{\rho \sigma}.
\end{equation}
As for the latter one, if we neglect the CP phase of the PMNS matrix in the following analysis, the LV terms can be expressed as
\begin{equation}\label{p1neq}
\begin{array}{ll}
\vspace{0.2cm}
P^{(1)}_{\alpha\rightarrow\beta}&={\displaystyle\sum_{i,j}}\hspace{0.05cm}\displaystyle{\sum_{\rho \neq \sigma}}\hspace{0.05cm} L\hspace{0.05cm} {\rm Im}\hspace{0.05cm}\left[(S^{(0)}_{\alpha \beta})^*\hspace{0.05cm}U_{\alpha i}\hspace{0.05cm} U_{\rho i} \hspace{0.05cm}\tau_{ij} \hspace{0.05cm} U_{\sigma j} \hspace{0.05cm}U_{\beta j}(\delta H_{\rho \sigma}+\delta H_{\sigma \rho})\right]\\
&={\displaystyle\sum_{i,j}}\hspace{0.05cm}\displaystyle{\sum_{\rho \neq \sigma}}\hspace{0.05cm}2\hspace{0.05cm} L\hspace{0.05cm} {\rm Im}\hspace{0.05cm}\left[(S^{(0)}_{\alpha \beta})^*\hspace{0.05cm}U_{\alpha i}\hspace{0.05cm} U_{\rho i} \hspace{0.05cm}\tau_{ij} \hspace{0.05cm}\hspace{0.05cm} U_{\sigma j} \hspace{0.05cm}U_{\beta j}\right] \hspace{0.05cm}{\rm Re}\hspace{0.05cm} \delta H_{\rho \sigma}.
\end{array}
\end{equation}
From the expressions in Eqs.~(\ref{p1eq}) and (\ref{p1neq}),
we notice that contributions from the mass term and LV terms can be factorized.
Therefore, we define the quantity $I_{\alpha \beta}^{\rho \sigma}$ as an indicator of the sensitivity to $\delta H_{\rho \sigma}$
in a certain oscillation channel $P_{\alpha\to\beta}$,
\begin{equation}\label{iterm}
I_{\alpha \beta}^{\rho \sigma}= {\displaystyle\sum_{i,j}}\hspace{0.05cm}2\hspace{0.05cm}L\hspace{0.05cm} {\rm Im}\hspace{0.05cm}\left[(S^{(0)}_{\alpha \beta})^*\hspace{0.05cm}U_{\alpha i}\hspace{0.05cm} U_{\rho i} \hspace{0.05cm}\tau_{ij} \hspace{0.05cm} U_{\sigma j} \hspace{0.05cm}U_{\beta j}\right].
\end{equation}
This indicator can be used to understand the distinct properties of different LV components $\delta H_{\rho \sigma}$ in the following numerical analysis.

\begin{figure}
\begin{minipage}[t]{0.49\textwidth}
\includegraphics[width=3in]{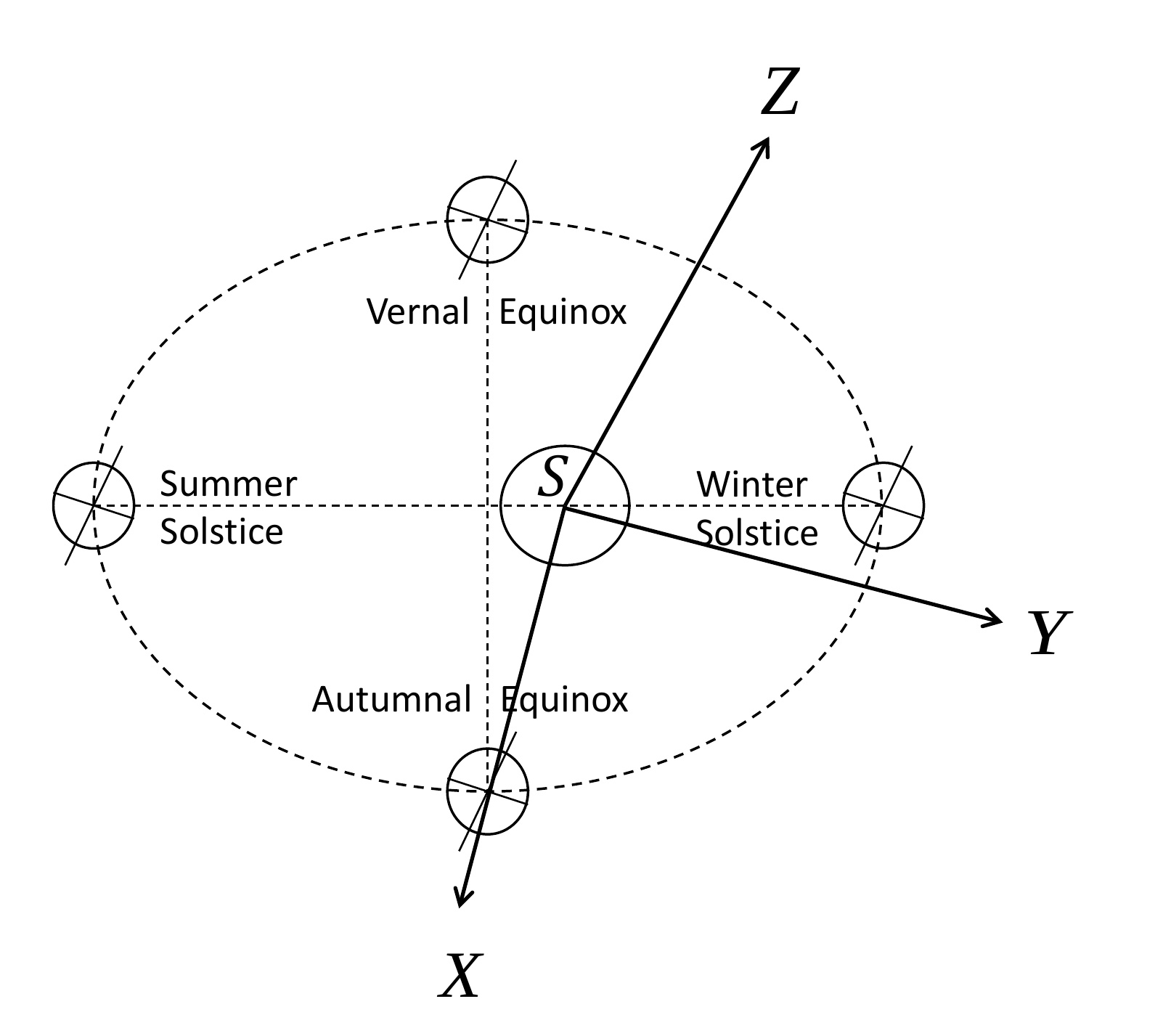}
\caption{The definition of the sun-centered system \cite{frame}.}
\label{sun}
\end{minipage}
\begin{minipage}[t]{0.49\textwidth}
\includegraphics[width=3in]{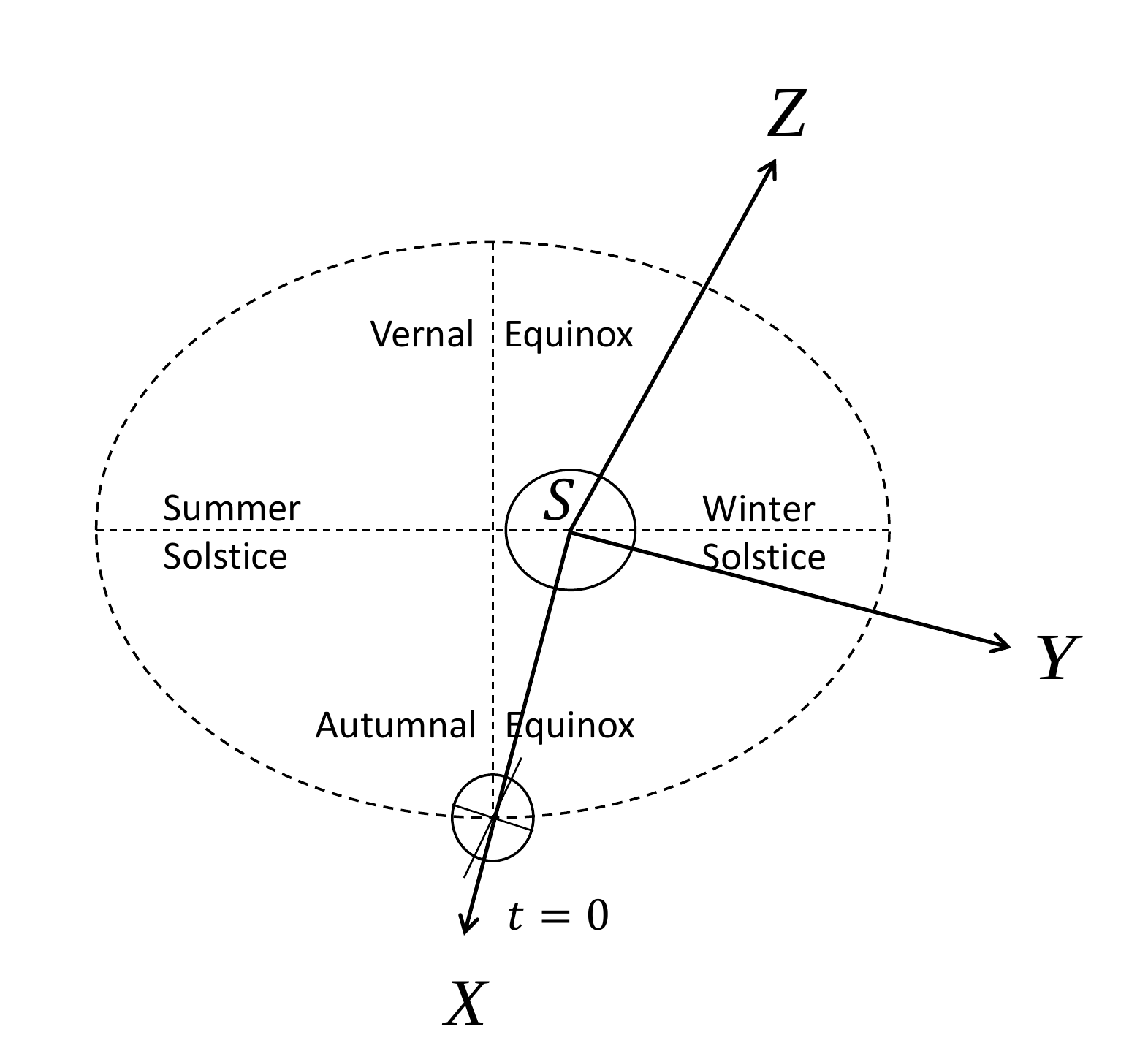}
\caption{The definition of $t=0$ in the sun-centered system.}
\label{time}
\end{minipage}
\end{figure}
In the above expressions of the oscillation probability, the most important property of the LV coefficients
is that $(a_{L})_\mu$ and $(c_{L})_{\mu \nu}$ are direction-dependent.
Thus a reference frame should be specified when an experiment
is going to report its LV results.
The sun can take on this responsibility, because it can be taken as an inertial frame to a good approximation \cite{frame}.
In the reference frame of the sun, the choice of the sun-centered system \cite{lsnd} is depicted in Fig.~\ref{sun} and defined in the following way:
\begin{itemize}
\item Point $S$ is the center of the sun;
\item The $Z$ axis has the same direction as the earth's rotational axis, so the $X$-$Y$ plane is parallel to the earth's equator;
\item The $X$ axis is parallel to the vector pointed from the sun to the autumnal equinox,
while the $Y$ axis completes the right-handed system.
\end{itemize}
\begin{figure}
\begin{minipage}[t]{0.49\textwidth}
\includegraphics[width=3in]{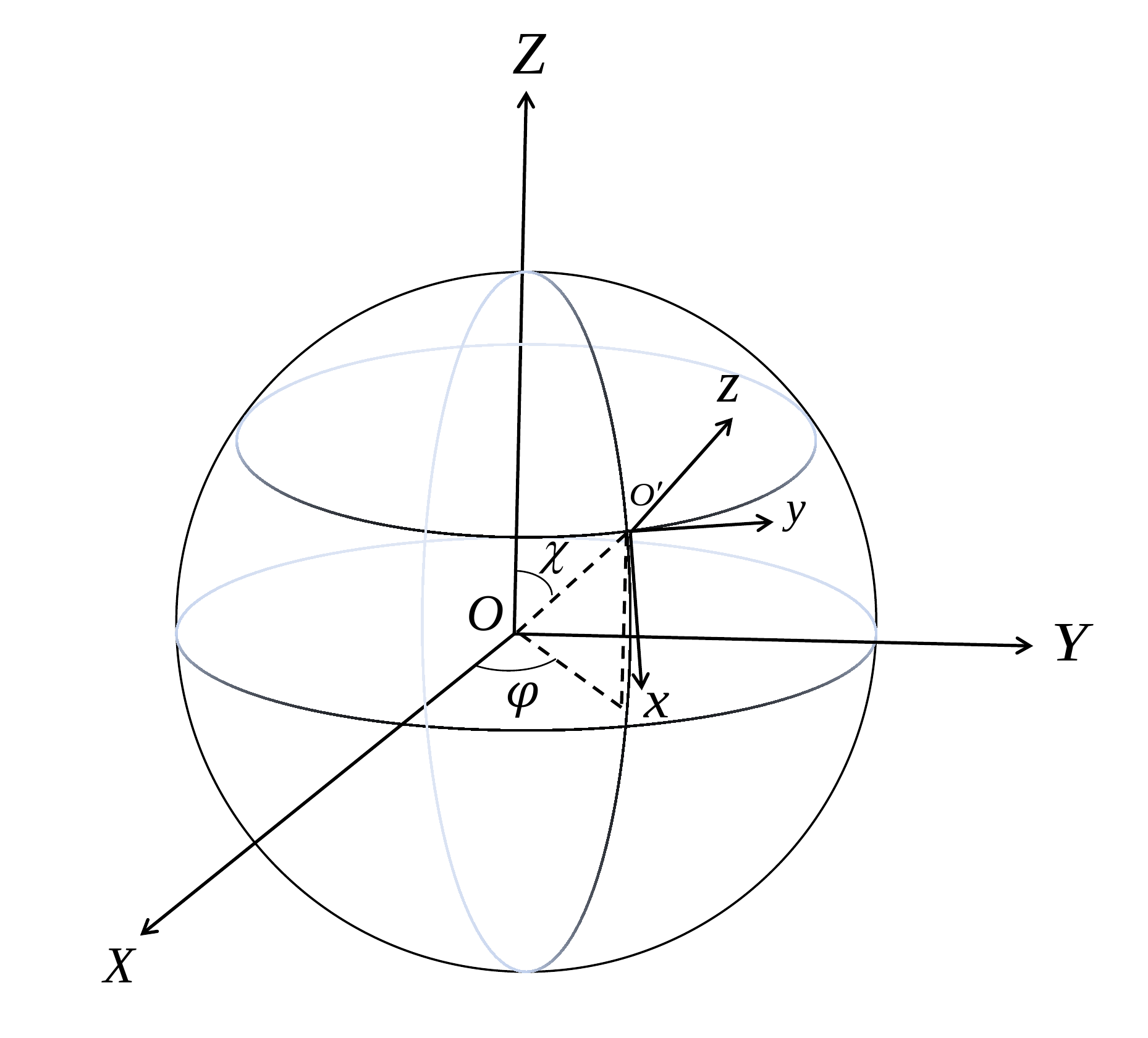}
\caption{The definition of the earth-centered system \cite{frame}.}
\label{earth}
\end{minipage}
\begin{minipage}[t]{0.49\textwidth}
\includegraphics[width=3in]{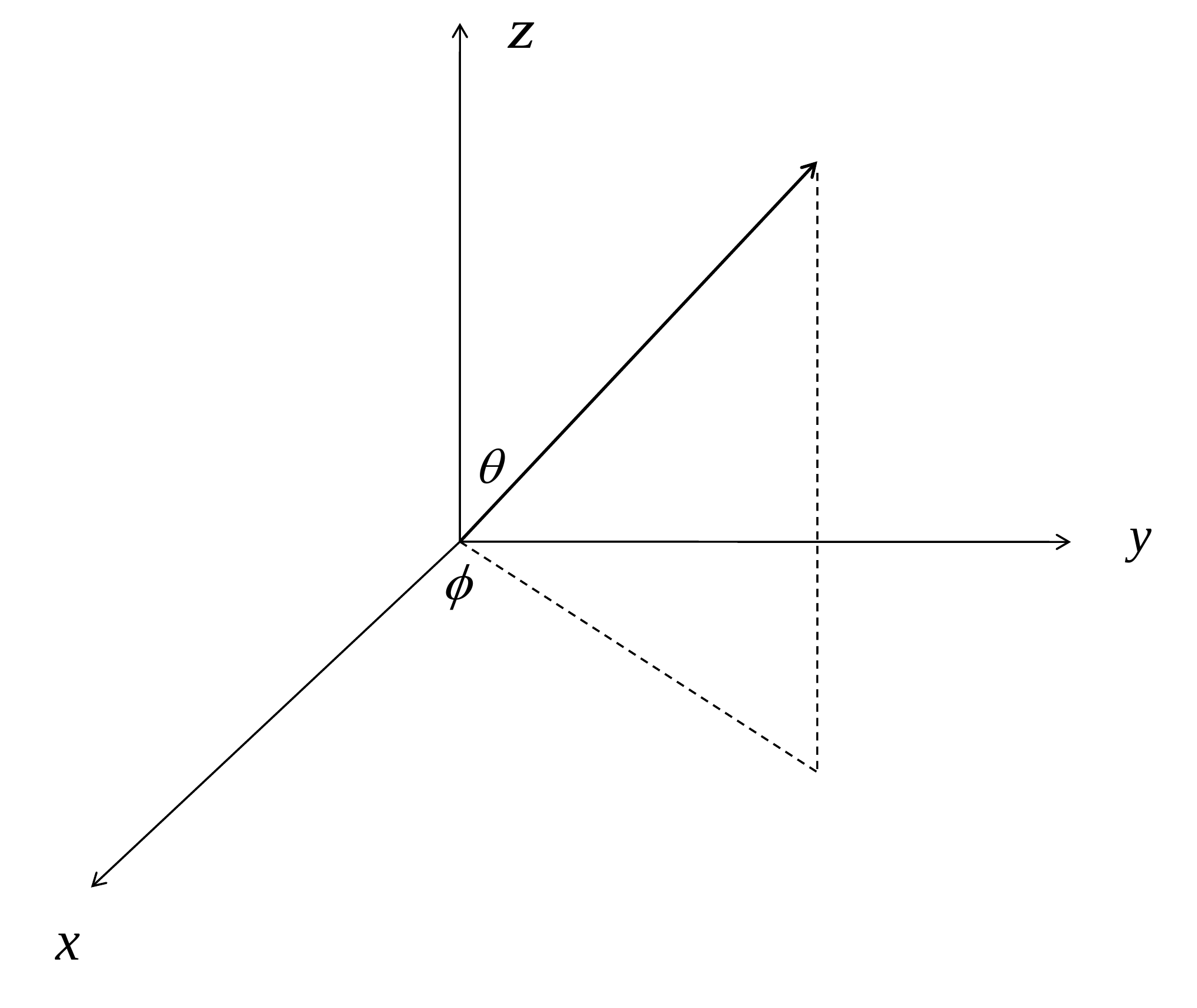}
\caption{The definition of the local coordinate system.}
\label{local}
\end{minipage}
\end{figure}
In a terrestrial experiment, the direction of neutrino propagation is described by
its components along the $X,Y,Z$ axes (i.e., $\hat N^{X}, \hat N^{Y}$ and $\hat N^{Z}$).
For convenience, the origin of the sun-centered system can be defined to be located in
the center of the earth $O$ due to the invariance of spatial translation, which is defined in Fig.~\ref{earth}.
In order to express $\hat N^{X}, \hat N^{Y}$ and $\hat N^{Z}$ in terms of local geographical information,
a local coordinate system $(x,y,z)$ is also introduced:
\begin{itemize}
\item point $O^\prime$ is the site of the neutrino source, and $\chi$ (i.e., the angle between $OO^\prime$ and $Z$ axis) denotes its colatitude;
\item the $z$ axis is defined to be upward;
\item the $x$ and $y$ axes point to the south and east, respectively.
\end{itemize}
In the local coordinate system depicted in Fig.~\ref{local}, the direction of neutrino propagation is parameterized by two angles, where
$\theta$ is the angle between the beam direction and $z$ axis and $\phi$ is the angle between the beam direction and $x$ axis.

Because the neutrino source and detector are fixed on the earth, the rotation of the earth will
induce a periodic change of the neutrino propagating directions relative to the standard reference frame,
with an angular frequency $\omega=2\pi/T_{\rm s}$.
Here $T_{\rm s}\simeq23 \hspace{0.1cm} {\rm h} \hspace{0.1cm}56\hspace{0.1cm} {\rm min}$ is the period of a sidereal day.
For this reason, a reference time origin should be specified.
Without loss of generality, we can set the local midnight when the earth arrives at the autumnal equinox to be $t=0$ (see Fig.~\ref{time}).
At this moment, the $x$-$z$ plane coincidences with the $X$-$Z$ plane, resulting in the coordinate transformation as
\begin{equation}
\begin{array}{lll}
\hat N^{X}&= & \cos{\chi} \sin{\theta} \cos{\phi}+ \sin{\chi} \cos{\theta},\\
\hat N^{Y}&= &\sin{\theta} \sin{\phi},\\
\hat N^{Z}&= &-\sin{\chi} \sin{\theta} \cos{\phi}+ \cos{\chi} \cos{\theta}.
\end{array}
\label{systrans}
\end{equation}
Therefore, the direction of neutrino propagation $\widehat{p^{\mu}}=(1,\ \widehat{p^{X}},\ \widehat{p^{Y}},\ \widehat{p^{Z}})$
is a periodic function of the time $t$
\begin{equation}
\begin{array}{lll}
\widehat{p^{X}}&=& \hat N^{X} \cos{\omega  t }- \hat N^{Y} \sin{\omega t },\\
\widehat{p^{Y}}&=& \hat N^{X} \cos{\omega  t }+ \hat N^{Y} \sin{\omega t },\\
\widehat{p^{Z}}&=& \hat N^{Z}.
\end{array}
\label{dirtrans}
\end{equation}
Accordingly, $\delta H_{\rho \sigma}$ can be decomposed as
\begin{equation}
\begin{array}{ll}
\delta H_{{\rho \sigma}}=&\mathcal C_{{\rho \sigma}}+(\mathcal A_s)_{\rho \sigma} \sin{\omega  t}+(\mathcal A_c)_{\rho \sigma} \cos{\omega  t}\\
&+(\mathcal B_s)_{\rho \sigma} \sin{2\omega  t}+ (\mathcal B_c)_{\rho \sigma} \cos{2\omega  t}.
\end{array}
\label{dh}
\end{equation}
Note that the above coefficients (i.e., $\mathcal C$, $\mathcal A_s$, $\mathcal A_c$,
$\mathcal B_s$ and $\mathcal B_c$) are linear combinations of the LV coefficients
$(a_{L})_\mu$ and $(c_{L})_{\mu \nu}$. Among them,
terms of $\mathcal A_s$, $\mathcal A_c$, $\mathcal B_s$ and $\mathcal B_c$ (written as $\mathcal {A/B}$ for short)
are time-dependent and can induce periodic variations for the oscillation probability.
On the other hand, $\mathcal C$ term can modify the absolute value of the oscillation probability with
unconventional energy and baseline dependence,
while the contributions of $\mathcal {A/B}$ cancel out in a full sidereal period.


\section{Numerical analysis}
In this section, we discuss two distinct LV effects in the medium baseline reactor antineutrino experiment. The first one is the spectral distortion
in the reactor antineutrino oscillation probability $P_{ee}$, and the second one is the sidereal modulation of the antineutrino event rate.
Previous experimental LV searches mainly focus on the signature of sidereal modulation as it can give definite evidence for LV and is independent
of the uncertainties of oscillation probabilities. However, as all the three mixing angles and two mass-squared differences are measured with high
precision, it is desirable to place compatible constraints on LV coefficients using the spectral distortion information.
Therefore, we employ both the phenomena to constrain LV coefficients.

In our numerical analysis, we take the JUNO experiment \cite{DYB2,JUNO} as a working example.
We simplify the reactor complexes in Yangjiang (labeled with ``1")
and Taishan (labeled with ``2") as two virtual reactors with equal baselines (i.e., 53 km).
Meanwhile, a 20 kton liquid scintillator detector is located in Jiangmen.
The antineutrinos coming from the two reactors can be written as
\begin{equation}
 \phi_1(E)=N_1\hspace{0.05cm} p_1(E)\ , \hspace{1cm}
 \phi_2(E)=N_2\hspace{0.05cm} p_2(E)\ ,
\end{equation}
where $N_1$ and $N_2$ are the total antineutrino events during the experimental period,
while $p_1(E)$ and $p_2(E)$ are their spectrum distributions.
For simplicity, we assume $p_1(E)=p_2(E)=p(E)$ and take the fuel composition as
$53.8\%$ ${}^{235}{\rm U}$, $7.8\%$ ${}^{238}{\rm U}$, $32.8\%$ ${}^{239}{\rm Pu}$ and $5.6\%$ ${}^{239}{\rm Pu}$.
The reactor antineutrino spectra are described by the following phenomenological expressions \cite{distribution},
\begin{equation}
\begin{array}{lll}
\vspace{0.3cm}
p_{235}(E)&=&\displaystyle\frac{1}{1.96}\hspace{0.05cm}\exp\hspace{0.05cm}[\hspace{0.05cm}3.2 - 3.1 E + 1.4 E^2 - 0.37  E^3 +0.045 E^4 -0.0021 E^5\hspace{0.05cm}]\ ,\\
\vspace{0.3cm}
p_{238}(E)&=&\displaystyle\frac{1}{2.53}\hspace{0.05cm}\exp\hspace{0.05cm}[\hspace{0.05cm}0.48 + 0.19 E - 0.13 E^2 - 0.0068 E^3 + 0.0022 E^4 -0.00015 E^5\hspace{0.05cm}]\ ,\\
\vspace{0.3cm}
p_{239}(E)&=&\displaystyle\frac{1}{1.52}\hspace{0.05cm}\exp\hspace{0.05cm}[\hspace{0.05cm}6.4 - 7.4 E + 3.5 E^2 - 0.88 E^3 + 0.10 E^4 -0.0046 E^5\hspace{0.05cm}]\ ,\\
p_{241}(E)&=&\displaystyle\frac{1}{1.88}\hspace{0.05cm}\exp\hspace{0.05cm}[\hspace{0.05cm}3.3 - 3.2 E + 1.4 E^2 - 0.37 E^3 + 0.043 E^4 -0.0019 E^5\hspace{0.05cm}]\ .
\end{array}
\end{equation}
As a consequence, $p(E)$ is a weighted average as
\begin{equation}
p(E)=0.538\hspace{0.05cm} p_{235}(E)+0.078\hspace{0.05cm} p_{238}(E)+ 0.328\hspace{0.05cm}p_{239}(E)+0.056\hspace{0.05cm}p_{241}(E)\ .
\end{equation}
Finally, the event number distribution can be obtained as a function of $E_{\rm o}$,
\begin{equation}\label{n0}
n^{(0)}(E_{\rm o}) = \int_{E_{\rm th}}^{\infty}\mathcal{N}\hspace{0.05cm}[N_1 + N_2]\hspace{0.05cm} p(E)\hspace{0.05cm} P_{ee}^{(0)}\hspace{0.05cm}
\sigma(E)\hspace{0.05cm}G(E_{\rm o},E)\hspace{0.05cm} dE\ . 
\end{equation}
Here $E_{\rm o}\simeq E_{e} + 1.3\,({\rm MeV})$, with $E_{e}$ being the energy of the outgoing positron in the inverse beta decay (IBD) process.
$G(E_{\rm o},E)$ is a Gaussian distribution for the energy resolution, with the standard deviation defined as
\begin{equation}
\rho \simeq 0.03 \sqrt{E_{\rm o}-0.8}\ .
\end{equation}
$\mathcal{N}$ is a normalization factor including the effects of reactor power, detection efficiency and the detector size.
With the approximation of equal baselines, the survival probability for $\bar \nu_e$
from the two different sources can be taken as the same one
\begin{equation}\label{}
  P_{ee}^{(0)}=1 - 4 s_{12}^2 c_{12}^2 c_{13}^4 \sin^2{ \frac{\Delta m_{21}^2L}{4E}} -
 4 s_{13}^2 c_{13}^2 c_{12}^2 \sin^2{ \frac{\Delta m_{31}^2L}{4E}} -
 4 s_{13}^2 c_{13}^2 s_{12}^2 \sin^2{\frac{\Delta m_{32}^2L}{4E}}\ ,
\end{equation}
where $s_{ij}=\sin\theta_{ij}$, $c_{ij}=\cos\theta_{ij}$ and $\Delta m_{ij}^2=m^{2}_{i}-m^{2}_{j}$
are the neutrino mixing angles defined in the standard parameterization \cite{PDG}
and mass squared differences, respectively.
$\sigma(E)$ is the leading-order IBD cross section \cite{cross}
\begin{equation}
\sigma(E)\simeq0.0952 (E-1.3) \sqrt{(E-1.3)^2-0.5^2}\times 10^{-42}\  {\rm cm}^2.
\end{equation}
As discussed in Ref.~\cite{JUNO}, about one hundred thousand IBD events are expected after a nominal running of six years. Therefore,
we normalize the total events in Eq.~({\ref{n0}}) to be one hundred thousand.

When LV effects are included, the energy distribution of the IBD events receives an extra contribution,
\begin{equation}\label{}
 n^{(1)}(E_{\rm o})=\int_{E_{\rm th}}^{\infty}\mathcal{N}\hspace{0.05cm} p(E)\hspace{0.05cm}\left[\hspace{0.05cm} N_1\hspace{0.05cm} (P_{ee}^{(1)})_1+ N_2\hspace{0.05cm}(P_{ee}^{(1)})_2\hspace{0.05cm}\right]\hspace{0.05cm}\sigma(E)\hspace{0.05cm}G(E_{\rm o},E)\hspace{0.05cm}dE\ .
\end{equation}
For simplicity, we define an effective LV-induced oscillation probability as
\begin{equation}
\begin{array}{ll}
\vspace{0.3cm}
P_{ee}^{(1){\rm eff}}&=\displaystyle \frac{1}{N_1+N_2}\hspace{0.05cm}\left[\hspace{0.05cm} N_1\hspace{0.05cm} (P_{ee}^{(1)})_1+ N_2\hspace{0.05cm}(P_{ee}^{(1)})_2\hspace{0.05cm}\right]\\
&={\displaystyle\sum_{i,j}}\hspace{0.05cm}{\displaystyle\sum_{\rho, \sigma}}\hspace{0.05cm}2\hspace{0.05cm}L\hspace{0.05cm} {\rm Im}\hspace{0.05cm}\left[\hspace{0.05cm}(S^{(0)}_{ee})^*\hspace{0.05cm} U_{e i}\hspace{0.05cm} U_{\rho i}^* \hspace{0.05cm}\tau_{ij}\hspace{0.05cm}\delta H_{\rho \sigma}^{\rm eff}\hspace{0.05cm} U_{\sigma j} \hspace{0.05cm}U_{e j}^*\hspace{0.05cm}\right]\ ,
\end{array}
\end{equation}
where
\begin{equation}\label{dheff}
  \delta H_{\rho \sigma}^{\rm eff}=\displaystyle \frac{1}{N_1 + N_2}\hspace{0.05cm}\left[\hspace{0.05cm}N_1\hspace{0.05cm}(\delta H_{\rho \sigma})_{1}+ N_2\hspace{0.05cm}(\delta H_{\rho \sigma})_{2}\hspace{0.05cm}\right]\ .
\end{equation}
Antineutrinos from two reactor sources have different propagation directions, so $(\delta H_{\rho \sigma})_1$ and $(\delta H_{\rho \sigma})_2$
can be expanded in the same way as Eq.~(\ref{dh}), with the information of the detector and reactor position \cite{pos}.
For LV coefficients $a_L^\mu$ and $c_L^{\mu \nu}$ with arbitrary directions,
the expansion coefficients for antineutrinos coming from Yangjiang are given as
\begin{equation}
\begin{array}{lll}
\vspace{0.2cm}
(\mathcal C_{\rho \sigma}^{*})_1&=&-a^{T}_{\rho \sigma}+0.81\  a^{Z}_{\rho \sigma}
+[\ 0.42\  c^{TT}_{\rho \sigma}-0.48\ c^{ZZ}_{\rho \sigma}\ ]\ E,\\
\vspace{0.2cm}
((\mathcal A_s)_{\rho \sigma}^{*})_1&=&-0.49\ a^{X}_{\rho \sigma}-0.33\ a^{Y}_{\rho \sigma}
-[\ 0.98\ c^{TX}_{\rho \sigma} +0.65\ c^{TY}_{\rho \sigma}-0.79\ c^{XZ}_{\rho \sigma}-
0.52\ c^{YZ}_{\rho \sigma}\ ]\ E,\\
\vspace{0.2cm}
((\mathcal A_c)_{\rho \sigma}^{*})_1&=&-0.33\  a^{X}_{\rho \sigma}+0.49\ a^{Y}_{\rho \sigma}
-[\ 0.65\ c^{TX}_{\rho \sigma} -0.98\ c^{TY}_{\rho \sigma}-0.52\ c^{XZ}_{\rho \sigma}+
0.79\ c^{YZ}_{\rho \sigma}\ ]\ E,\\
\vspace{0.2cm}
((\mathcal B_s)_{\rho \sigma}^{*})_1&=&-0.16\ c_{\rho \sigma}^{XX}+0.16\ c_{\rho \sigma}^{YY}
+0.13\ c_{\rho \sigma}^{XY},\\
\vspace{0.2cm}
((\mathcal B_c)_{\rho \sigma}^*)_1&=&-0.067\ c_{\rho \sigma}^{XX}+0.067\ c_{\rho \sigma}^{YY}
+0.32\ c_{\rho \sigma}^{XY}.
\label{yj}
\end{array}
\end{equation}
while the expansion coefficients for antineutrinos coming from Taishan are given as
\begin{equation}
\begin{array}{lll}
\vspace{0.2cm}
(\mathcal C_{\rho \sigma}^{*})_2&=&-a^{T}_{\rho \sigma}+0.41\  a^{Z}_{\rho \sigma}
-[\ 0.59\  c^{TT}_{\rho \sigma}-0.25\ c^{ZZ}_{\rho \sigma}\ ]\ E,\\
\vspace{0.2cm}
((\mathcal A_s)_{\rho \sigma}^{*})_2&=&0.90\ a^{X}_{\rho \sigma}-0.17\ a^{Y}_{\rho \sigma}
+[\ 1.8\ c^{TX}_{\rho \sigma} -0.34\ c^{TY}_{\rho \sigma}-0.73\ c^{XZ}_{\rho \sigma}+
0.14\ c^{YZ}_{\rho \sigma}\ ]\ E,\\
\vspace{0.2cm}
((\mathcal A_c)_{\rho \sigma}^{*})_2&=&-0.17\  a^{X}_{\rho \sigma}-0.90\ a^{Y}_{\rho \sigma}
-[\ 0.34\ c^{TX}_{\rho \sigma} +1.8\ c^{TY}_{\rho \sigma}-0.14\ c^{XZ}_{\rho \sigma}-
0.73\ c^{YZ}_{\rho \sigma}\ ]\ E,\\
\vspace{0.2cm}
((\mathcal B_s)_{\rho \sigma}^{*})_2&=&0.15\ c_{\rho \sigma}^{XX}-0.15\ c_{\rho \sigma}^{YY}
+0.77\ c_{\rho \sigma}^{XY},\\
\vspace{0.2cm}
((\mathcal B_c)_{\rho \sigma}^*)_2&=&-0.39\ c_{\rho \sigma}^{XX}+0.39\ c_{\rho \sigma}^{YY}
-0.30\ c_{\rho \sigma}^{XY}.
\label{ts}
\end{array}
\end{equation}
Here the subscript ``$_L$" for $a_L$ and $c_L$ is omitted for simplicity.

The energy range from 1.8 to 8 $\hspace{0.1cm} {\rm MeV}$ is divided into 200 equal-size bins.
For each bin with the label $i$, the expected event number can be calculated as
\begin{equation}\label{}
 n_i^{(0)}=\int_{E_i-\Delta E}^{E_i+\Delta E} n^{(0)}(E_{\rm o})\hspace{0.1cm}dE_{\rm o}\ ,
\end{equation}
where $\Delta E\simeq0.031\hspace{0.1cm} {\rm MeV}$ and
$E_i=[\hspace{0.05cm}1.8+(i-0.5)\hspace{0.05cm}\Delta E\hspace{0.05cm}]\hspace{0.1cm} {\rm MeV}$.
In this calculation, the oscillation parameters are taken as the central values of the recent global-fit analysis \cite{fit},
\begin{equation}\label{}
\begin{array}{ll}
  \sin^2{\theta_{12}}=0.323\ , &\hspace{1cm} \sin^2{\theta_{13}}=0.0234\ ;\\
  \Delta m_{21}^2=7.60\times 10^{-5}\ \rm{eV^2}\ , &\hspace{1cm} \Delta m_{31}^2=2.48\times 10^{-3}\ \rm{eV^2}\ .
  \end{array}
\end{equation}
Furthermore, the event numbers from the LV terms are calculated as
\begin{equation}\label{}
 n_i^{(1)}=\int_{E_i-\Delta E}^{E_i+\Delta E} n^{(1)}(E_{\rm o})\hspace{0.1cm}dE_{\rm o}\ .
\end{equation}
Because $\delta H_{\rho \sigma}^{\rm eff}$ has an identical form as Eq.~(\ref{dh}), we can define
the effective coefficients $\mathcal C_{\rho \sigma}^{\rm eff}$ and $\mathcal {A/B}_{\rho \sigma}^{\rm eff}$
as the weighted averages of $ (\mathcal C_{\rho \sigma})_{1/2}$ and $(\mathcal {A/B}_{\rho \sigma})_{1/2}$.

$\mathcal C_{\rho \sigma}^{\rm eff}$ can modify the energy dependence of the oscillation probability.
\begin{figure}
\begin{center}
\begin{tabular}{cc}
\includegraphics*[width=0.45\textwidth]{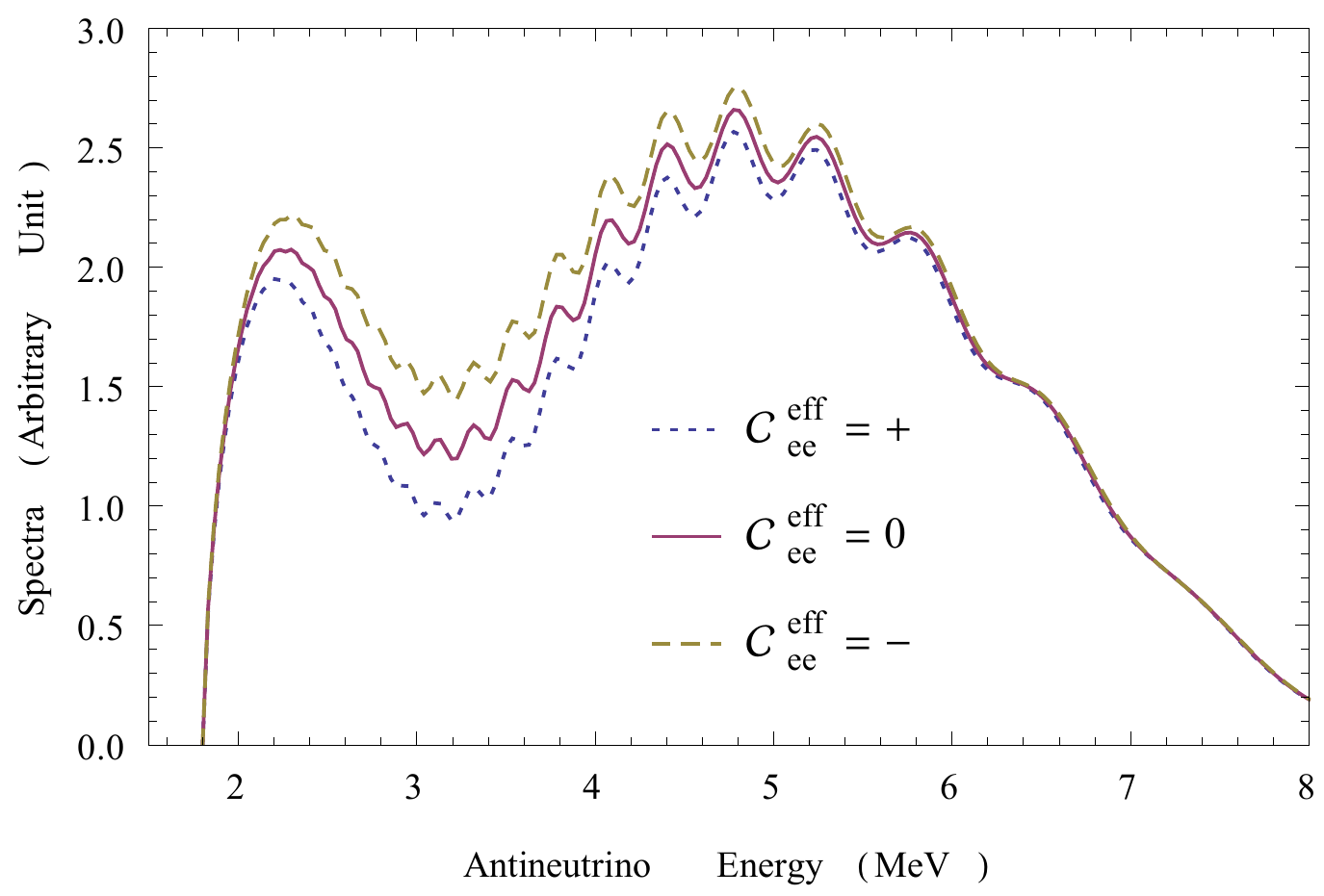}
&
\includegraphics*[width=0.45\textwidth]{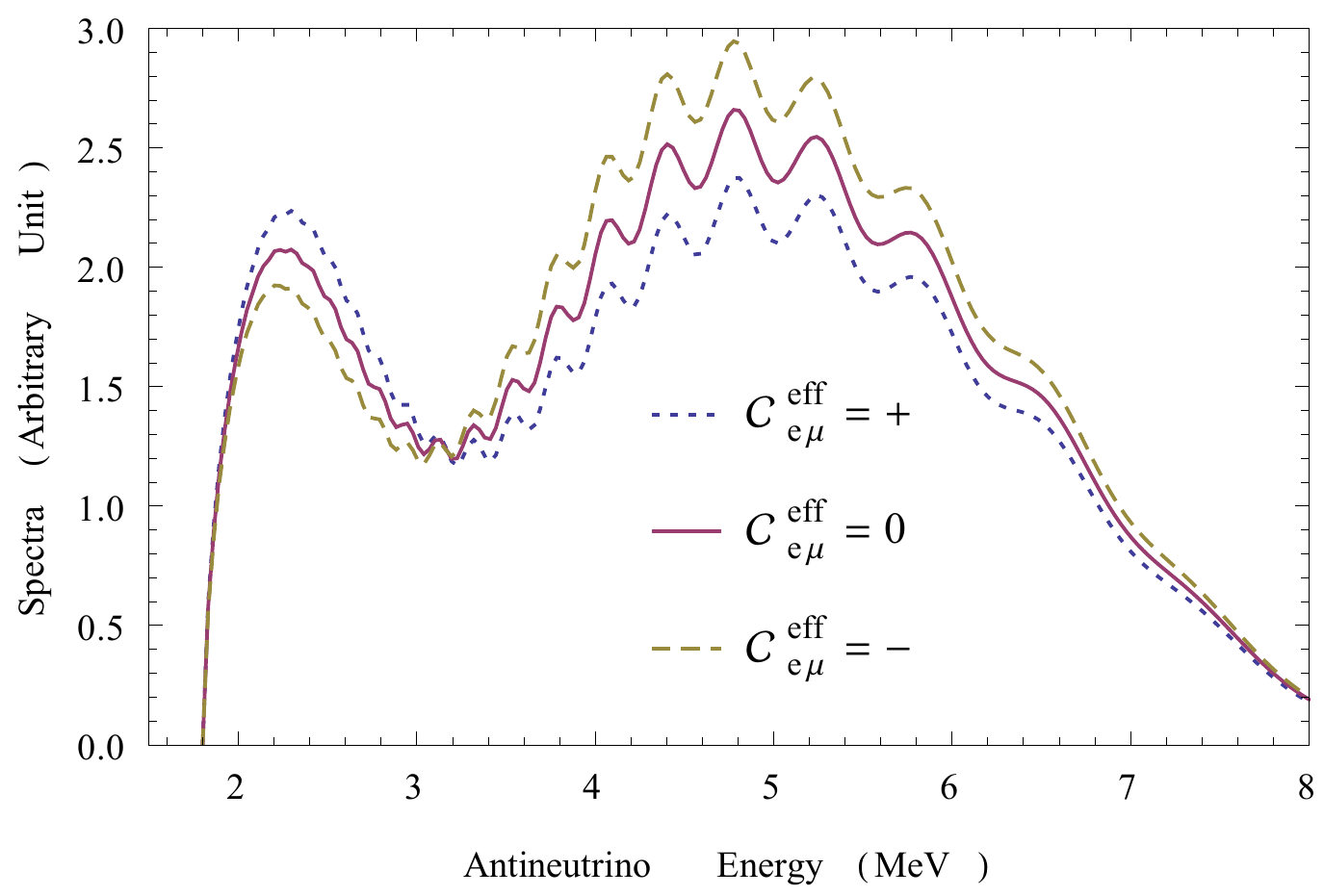}
\\
\includegraphics*[width=0.45\textwidth]{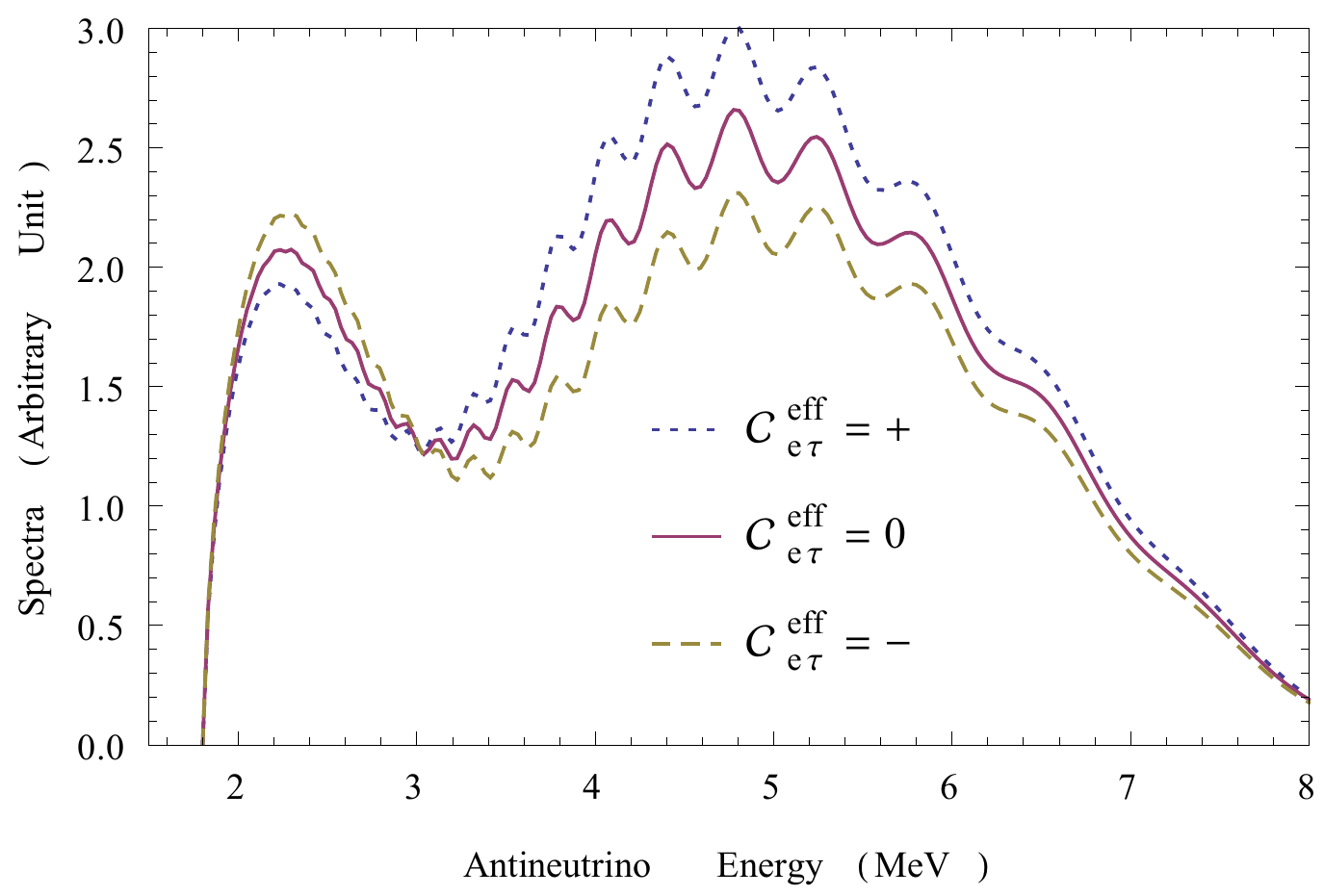}
&
\includegraphics*[width=0.45\textwidth]{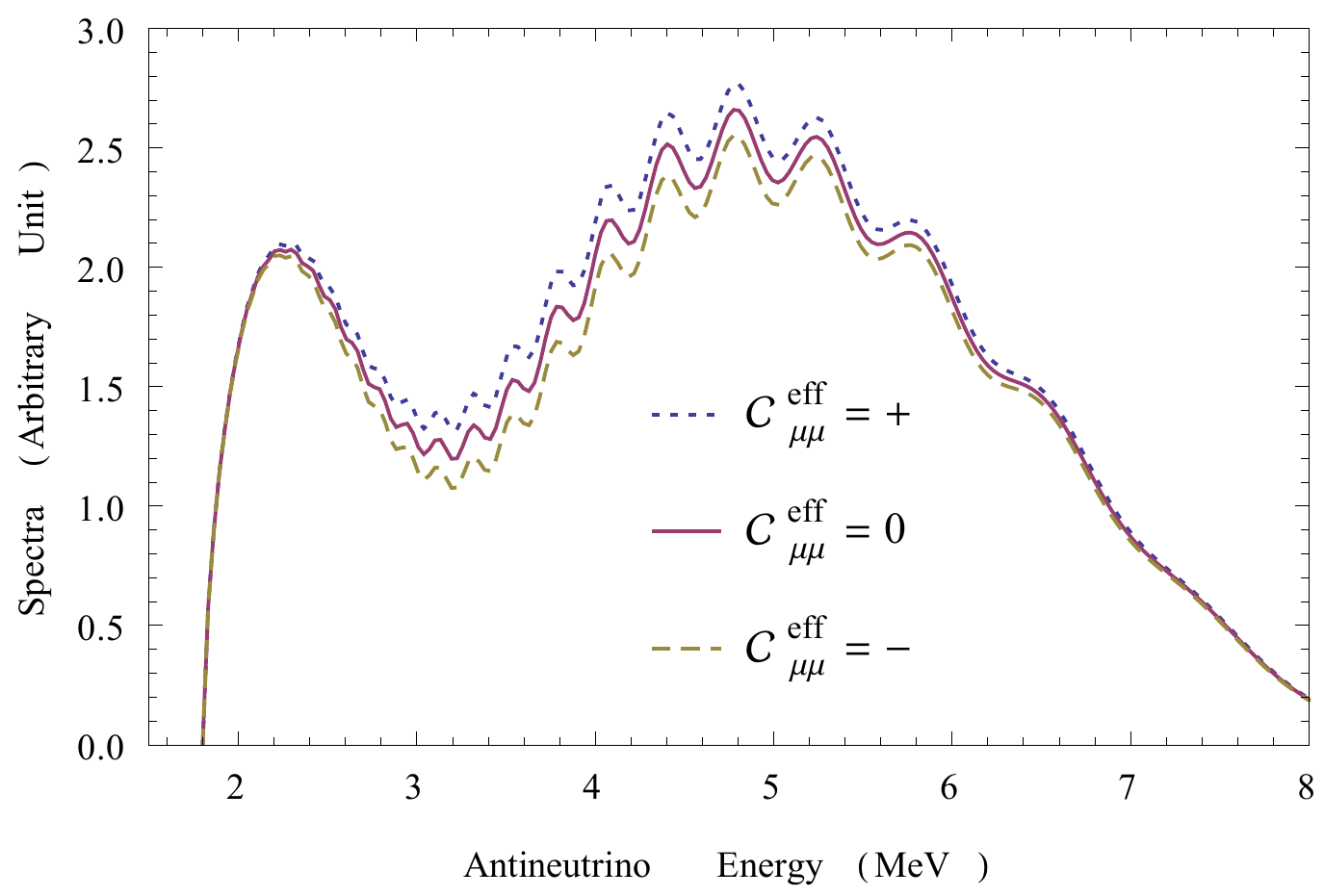}
\\
\includegraphics*[width=0.45\textwidth]{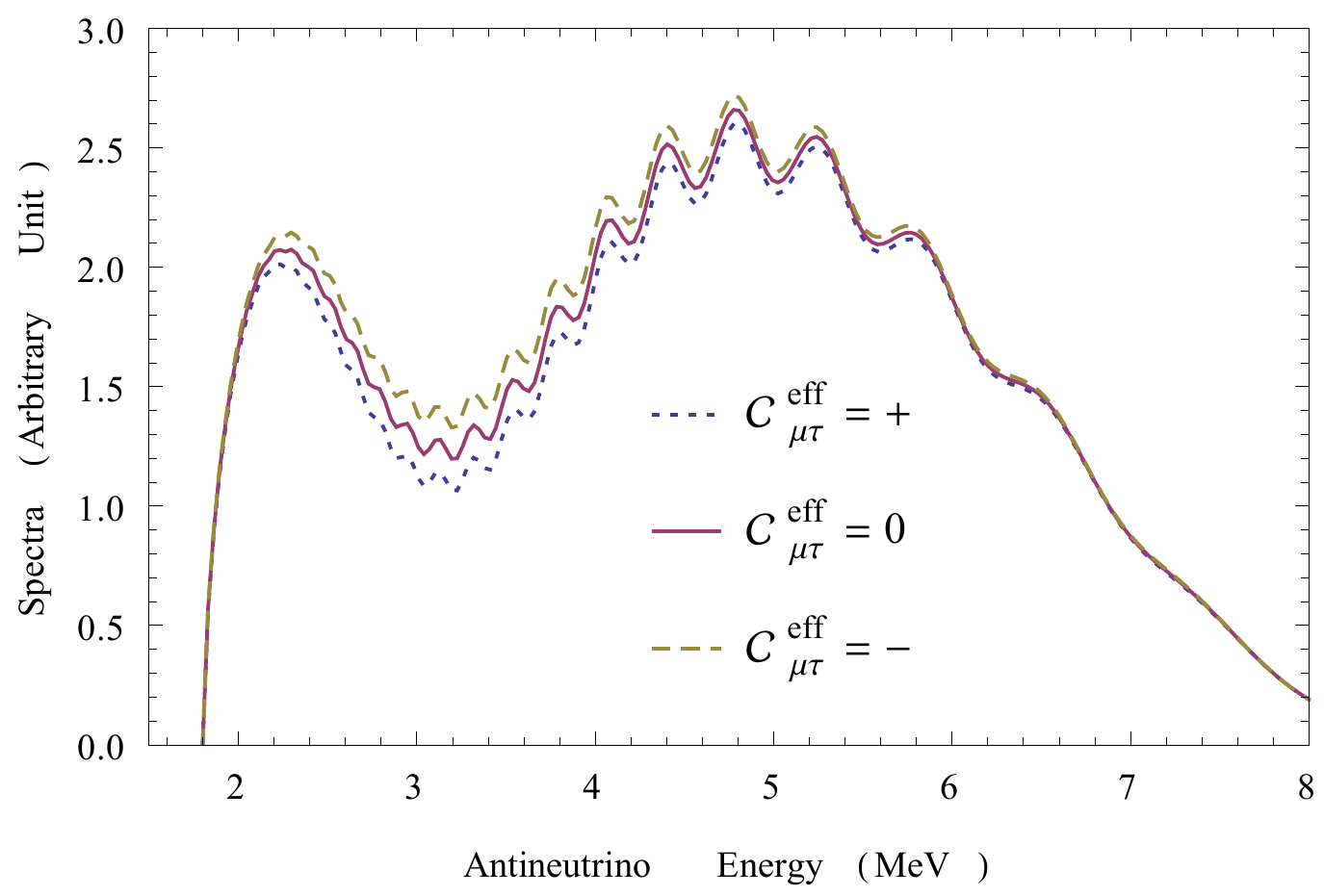}
&
\includegraphics*[width=0.45\textwidth]{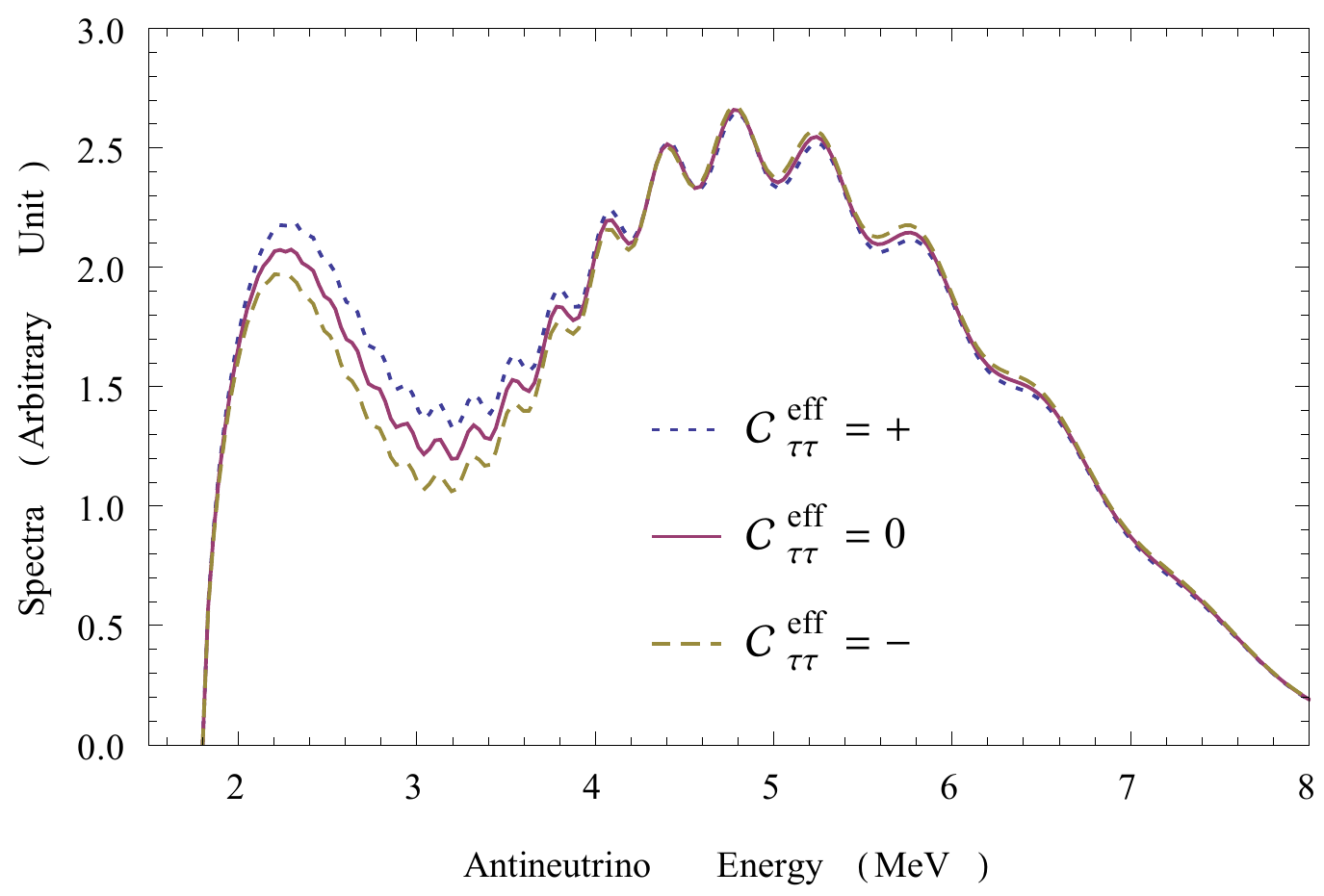}
\end{tabular}
\end{center}
\caption{Spectral distortion effects for the six effective LV coefficients. Their magnitudes are taken as
$(\pm5.0\times10^{-22})\, {\rm GeV}$ and zero for comparison.}
\label{spec}
\end{figure}
Fig.~(\ref{spec}) shows the spectral distortion effects for the six effective LV coefficients $\mathcal C_{\rho \sigma}^{\rm eff}$.
Their magnitudes are taken as
$(\pm5.0\times10^{-22})\, {\rm GeV}$ for comparison. The standard case without LV is also illustrated.
Therefore, we can constrain the magnitude of $\mathcal C_{\rho \sigma}^{\rm eff}$ by comparing the observed spectrum with the expected one.
There are many $\mathcal C_{\rho \sigma}^{\rm eff}$ components and they may be cancelled with each other when combined together.
For definiteness, we shall take only one non-zero coefficient at one time in the analysis.

\begin{figure}
\begin{center}
\begin{tabular}{c}
\includegraphics*[width=0.6\textwidth]{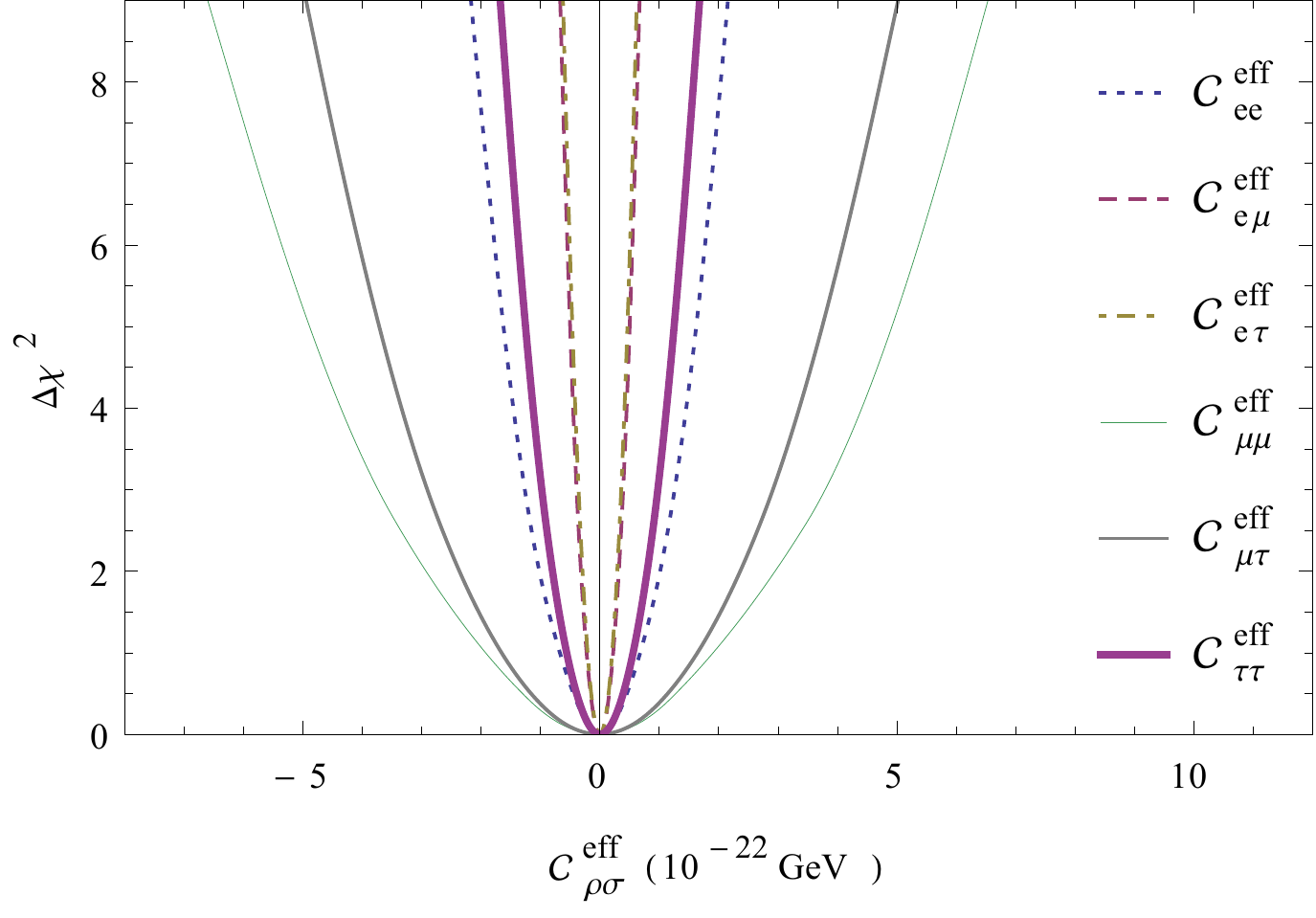}
\\
\includegraphics*[width=0.6\textwidth]{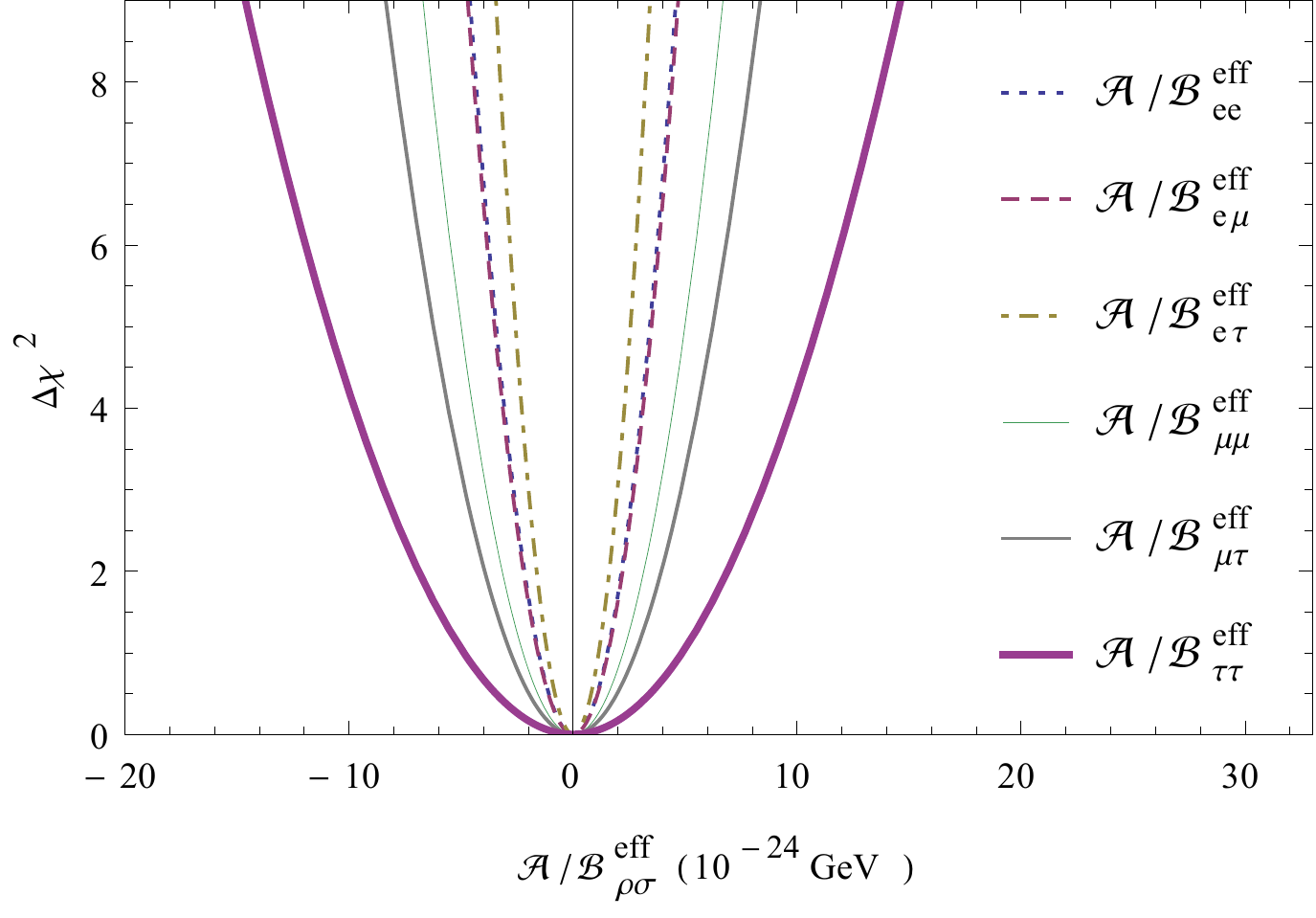}
\end{tabular}
\end{center}
\caption{The JUNO sensitivity for the effective LV coefficients from the spectral distortion (upper panel) and sidereal variation (lower panel) effects.}
\label{chi2}
\end{figure}
To quantify the significance of LV effects in the spectral distortion, a $\chi^2$ function is defined as follows:
\begin{equation}\label{ch2spec}
\chi^2=\sum_i \frac{\left[n_i^{(0)}(s^{2}_{12})-\left(n_i^{(0)\prime}(s_{12}^{\prime 2})+n_i^{(1)}\right)(1+\epsilon)\right]^2}{n_i^{(0)}(s^{2}_{12})+\left(0.01 n_i^{(0)}(s^{2}_{12})\right)^2}
+\frac{\epsilon^2}{0.05^2}+\frac{(s_{12}^{\prime 2}-s_{12}^2)^2}{0.016^2}\,,
\end{equation}
where two pull-parameters are included. $\epsilon$ is a normalization factor with an uncertainty of 0.05,
$s_{12}^{\prime 2}$ is the oscillation parameter that affects the analysis most.
The uncertainty of $s_{12}^{2}$ is also taken from Ref.~\cite{fit}. We also take a spectral uncertainty of the $1\%$ level,
which is assumed uncorrelated among different energy bins.
For each $\mathcal C_{\rho \sigma}^{\rm eff}$, we can obtain a one-dimensional $\Delta \chi^2$ function
after marginalizing the pull-parameters $\epsilon$ and $s_{12}^{\prime 2}$.
The distributions of $\Delta \chi^2$ as functions of the six effective LV coefficients $\mathcal C_{\rho \sigma}^{\rm eff}$
are shown in the upper panel of Fig.~(\ref{chi2}).
The upper limits at the $95\%$ confidence level for these LV coefficients are listed in the first row of Table~\ref{limit}.
The relative differences in the power of constraining these six coefficients can be understood by the quantity defined in
Eq.~({\ref{iterm}}), which indicates that the flavor indices with larger $I_{\alpha \beta}$ will get more severe constraints.

\begin{table}
\begin{tabular}{|p{2.5cm}<{\centering}|p{2cm}<{\centering}|p{2cm}<{\centering}|p{2cm}<{\centering}
|p{2cm}<{\centering}|p{2cm}<{\centering}|p{2cm}<{\centering}|}
\hline
   &$\mathcal C_{ e e}^{{\rm eff}}$ &$ \hspace{0.1cm}\mathcal C_{ e \mu}^{{\rm eff}}$  &$\hspace{0.1cm}\mathcal C_{ e  \tau}^{{\rm eff}}$
   &$\mathcal C_{\mu \mu}^{{\rm eff}}$ &$\hspace{0.1cm}\mathcal C_{ \mu  \tau}^{{\rm eff}}$  &$\mathcal C_{ \tau  \tau}^{{\rm eff}}$ \\
         \hline
$10^{-22}\ {\rm GeV}$  & 1.5  &   0.5   & 0.4 & 4.4   &  3.4   & 1.1  \\
\hline

\hline
   &$\mathcal {A/B}_{e e}^{{\rm eff}}$ &$\hspace{0.1cm}\mathcal {A/B}_{e \mu}^{{\rm eff}}$  &$\hspace{0.1cm}\mathcal {A/B}_{e \tau}^{{\rm eff}}$
   &$\mathcal {A/B}_{\mu \mu}^{{\rm eff}}$ &$\hspace{0.1cm}\mathcal {A/B}_{\mu \tau}^{{\rm eff}}$  &$ \mathcal {A/B}_{\tau \tau}^{{\rm eff}}$ \\
         \hline
$10^{-24}\ {\rm GeV}$  & 3.7 &   3.7   &  2.7  & 5.4   & 6.6   &  11.6    \\
\hline
\end{tabular}
\caption{The JUNO sensitivity at the $95\%$ confidence level for the effective LV coefficients $\mathcal C$, $\mathcal {A/B}$ in $\delta H_{\rho \sigma}^{{\rm eff}}$.}
\label{limit}
\end{table}

Next we shall discuss the constraints on the effective coefficients $\mathcal {A/B}_{\rho \sigma}^{\rm eff}$
from the sidereal variation of IBD event rates.
For this purpose, a sidereal day is divided into 24 bins.
Each bin $j$ is expected to have $n_j^{(0)}\simeq100000/24\simeq4167$ IBD events in the whole energy range from 1.8 to 8 $\rm {MeV}$.
Similar to Eq.~(\ref{ch2spec}), another $\chi^2$ is defined as
\begin{equation}
\chi^2=\sum_j \frac{n_j^{(1)2}}{n_j^{(0)}+\left(0.01 n_j^{(0)}\right)^2}\ .
\end{equation}
Here $n_j^{(1)}$ is calculated through the following expression,
\begin{equation}
n_j^{(1)}= \int_{T_{j-1}}^{T_j}\left[\int_{E_{\rm th}}^{\infty}n^{(1)}(E_{\rm o})\hspace{0.05cm}dE_{\rm o}\right]dT,
\end{equation}
with $T_j=j \hspace{0.1cm}{\rm h}$.
Notice that a normalization factor like $\epsilon$ in Eq.~(\ref{ch2spec}) has a negligible effect in the analysis.
Thus we only consider the statistical uncertainty of $n_j^{(0)}$ and the time-dependent uncorrelated uncertainty of the $1\%$ level.
The $\Delta \chi^2$ as functions of the effective LV coefficients $\mathcal {A/B}_{\rho \sigma}^{\rm eff}$ are shown in the lower panel of Fig.~(\ref{chi2}).
The $95\%$ upper limits for these LV coefficients are listed in the second row of Table~\ref{limit},
where the $\tau\tau$ component gets the worst sensitivity, but the $e\tau$ coefficient turns out to be the most severely constrained parameters.
The order of magnitude for these coefficients is $10^{-24}$ GeV, much smaller than that for $\mathcal C_{\rho \sigma}^{\rm eff}$ (i.e., $10^{-22}$ GeV).
This is because the uncertainties of the spectrum and neutrino oscillation parameters do not enter the sidereal variation of IBD events.

It is straightforward to transfer the limits for $\mathcal C_{\rho \sigma}^{\rm eff}$ and $\mathcal {A/B}_{\rho \sigma}^{\rm eff}$
to that for each space component of the physical parameters $a_{L}^\mu$ and $c_{L}^{\mu \nu}$,
with the help of the relations given by Eqs.~(\ref{yj}) and (\ref{ts}). 
However, the degrees of freedom in $a_{L}^\mu$ and $c_{L}^{\mu \nu}$ are much larger than those in the effective LV coefficients.
In order to obtain the independent constraints, it is more reasonable to use the effective coefficients.
On the other hand, one should derive the limits for $a_{L}^\mu$ and $c_{L}^{\mu \nu}$
when comparing and combining the limits from different oscillation experiments.

\section{Conclusion}
In this work, we have presented the sensitivity study for the Lorentz and CPT violation in the medium baseline reactor antineutrino experiment.
Taking the JUNO experiment as an illustration, we calculate the sensitivity for constraining the LV coefficients using both the spectral distortion
and sidereal variation effects. The time-independent spectral distortion can constrain the effective LV coefficients $\mathcal C_{\rho \sigma}^{\rm eff}$
to the level of $10^{-22}$ GeV, and using the sidereal variation effect one can test LV with a precision of $10^{-24}$ GeV.
Considering the flavor indices of the LV coefficients, we have the best constraints for the $e\tau$ components, but the worst ones for the $\mu\mu$ or
$\tau\tau$ components. The JUNO sensitivity is at least two orders of magnitude better than
the achieved limits in the reactor antineutrino experiment \cite{chooz} due to the longer baseline and much better energy resolution.

Although suppressed by the factor of the order of the electro-week scale divided by the Planck scale (i.e, $10^{-17}$), the low-energy
neutrino oscillation phenomena provide us an accessible tool to test Planck scale physics. Comparing and combining with other probes of Lorentz and CPT
violation, including the high energy tests with cosmic rays and astrophysical neutrinos, we may be able to reveal the hidden information behind the Planck scale
and obtain the possible clue of Quantum Gravity \cite{ssb}.

\section*{Acknowledgements}
We are indebted to Z.~Z.~Xing for his continuous encouragement and useful discussions. We are also grateful to J.~Cao and L.~Zhan for providing
the information of the JUNO detector and reactor position. This work was supported in part by the National Natural Science Foundation of China
under Grant Nos. 11135009, 11305193, and in part by the Strategic Priority Research Program of the Chinese Academy of Sciences under Grant No. XDA10010100.


\begin{thebibliography}{99}
\bibitem{pos}
The information of the detector and reactor position is from Drs. Jun Cao and Liang Zhan.
\end{thebibliography}
\end{document}